\pdfoutput=1 
\documentclass[10pt,conference]{IEEEtran}

\usepackage{code}
\usepackage{graphicx}
\usepackage{balance}
\usepackage{multirow}
\usepackage{multicol}
\usepackage{listings}
\usepackage{booktabs}
\usepackage{subcaption}
\usepackage[normalem]{ulem}
\usepackage{hyperref}
\usepackage{cleveref}
\usepackage{xspace}
\usepackage{xcolor}
\usepackage{svg}
\definecolor{deepred}{rgb}{0.6,0,0}
\definecolor{dkgreen}{rgb}{0,0.6,0}
\definecolor{lightgray}{rgb}{0.95,0.95,0.95}
\definecolor{circleblue}{HTML}{4472C4} 

\usepackage{circledsteps}
\newcommand*{\circlednumber}[1]{%
  \pgfkeys{/csteps/inner color=white}%
  \pgfkeys{/csteps/fill color=circleblue}%
  \Circled[]{\textbf{\textsf{#1}}}}

\makeatletter
\newcommand{\linebreakand}{%
  \end{@IEEEauthorhalign}
  \hfill\mbox{}\par
  \mbox{}\hfill\begin{@IEEEauthorhalign}
}
\makeatother

\begin{document}

\author{
\IEEEauthorblockN{Kush Jain}
\IEEEauthorblockA{\textit{Carnegie Mellon University} \\ 
United States \\
kdjain@andrew.cmu.edu}
\and
\IEEEauthorblockN{Kiran Kate}
\IEEEauthorblockA{\textit{IBM Research} \\ 
United States \\
kakate@us.ibm.com}
\and
\IEEEauthorblockN{Jason Tsay}
\IEEEauthorblockA{\textit{IBM Research} \\ 
United States \\
Jason.Tsay@ibm.com}
\linebreakand 
\IEEEauthorblockN{Claire Le Goues}
\IEEEauthorblockA{\textit{Carnegie Mellon University} \\ 
United States \\
clegoues@cs.cmu.edu}
\and
\IEEEauthorblockN{Martin Hirzel}
\IEEEauthorblockA{\textit{IBM Research} \\ 
United States \\
hirzel@us.ibm.com}
}

\title{Improving Examples in Web API Specifications using Iterated-Calls In-Context Learning}



\newcommand{\mr}[2]{\multirow{#1}{*}{#2}}
\newcommand{\mc}[3]{\multicolumn{#1}{#2}{#3}}

\newcommand{\kj}[1]{\textcolor{olive}{\textbf{\small[Kush: #1]}}}
\newcommand{\todo}[1]{\textcolor{red}{todo: #1}}

\newcommand{\technique}{\textsc{IcIcl}\xspace}
\newcommand{\techniquecap}{\technique}

\lstdefinelanguage{yaml}{
  basicstyle=\footnotesize\ttfamily,        
  breaklines=true,                   
  frame=single,                      
  xleftmargin=2em,                  
  linewidth=\linewidth,              
  morestring=[b]',
  morestring=[b]",
  morecomment=[l]{\#},
  morekeywords={true,false,null,y,n, Model, Input, Output},
  keywordstyle=\color{blue}\bfseries,  
  stringstyle=\color{deepred},         
  commentstyle=\color{dkgreen},        
  rulecolor=\color{black},             
  backgroundcolor=\color{lightgray},   
  showspaces=false,                    
  showstringspaces=false,              
  showtabs=false,                      
  tabsize=2,                           
}

\lstdefinelanguage{kotlin}{
    basicstyle=\ttfamily\footnotesize,
    frame=single,                      
    xleftmargin=2em,                  
    keywordstyle=\color{blue},
    commentstyle=\color{dkgreen},
    stringstyle=\color{deepred},
    breaklines=true,
    breakatwhitespace=true,
    tabsize=2,
    keywords={if, when, return, true, as, add, apply},
    comment=[l]{//},
    morestring=[b]",   
    backgroundcolor=\color{lightgray},   
    rulecolor=\color{black},             
}

\lstdefinestyle{conversation}{
  basicstyle=\ttfamily\footnotesize,
  breaklines=true,
  columns=fullflexible,
  commentstyle=\color{black},
  morecomment=[l]{[},
  morecomment=[l]{\}},
  morekeywords={example, user, system, slots, context, state, Model, Input, Output, Example, Conversation},
  keywordstyle=\color{blue}\bfseries,
  frame=single,
  backgroundcolor=\color{lightgray},
}

\maketitle

\begin{abstract}
Examples in web API specifications can be essential for API testing,
API understanding, and even building chat-bots for APIs.
Unfortunately, most API specifications lack human-written examples.
This paper introduces a novel technique for generating examples for
web API specifications.
We start from in-context learning (\textsc{Icl}): given an API
parameter, use a prompt context containing a few examples from other
similar API parameters to call a model to generate new examples.
However, while \textsc{Icl} tends to generate correct examples, those
lack diversity, which is also important for most downstream tasks.
Therefore, we extend the technique to
iterated-calls \textsc{Icl}~(\technique): use a few different prompt
contexts, each containing a few examples, to iteratively call the
model with each context.
Our intrinsic evaluation demonstrates that \technique improves both
correctness and diversity of generated examples.
More importantly, our extrinsic evaluation demonstrates that those
generated examples significantly improve the performance of downstream
tasks of testing, understanding, and chat-bots for APIs.

\end{abstract}

\section{Introduction}

Web Application Programming Interfaces (APIs) enable systems to communicate
across a network~\cite{Jacobson2011APIsAS, RichardsonRESTAPIBook}.
REpresentational State Transfer (REST) APIs have become the de facto 
standard for modern web applications~\cite{restapithesis}. This style enables clients and services to exchange information
over HTTP.  Large companies like Google, Amazon, and Apple expose services through REST APIs, including large enterprise services like 
Google Drive and Apple Authentication,
as well as simpler services like REST Countries,\footnote{\url{https://restcountries.com}} for querying information about a country.

REST APIs are commonly described using OpenAPI specifications~\cite{OpenAPI}: one survey of
communication service providers found that 
73\% of companies and 75\% of suppliers use OpenAPI to describe their APIs.\footnote{\url{https://inform.tmforum.org/features-and-opinion/the-status-of-open-api-adoption/}}
OpenAPI specifications formalize the contract between API developer and API user, describing the structure 
of API requests and responses. Tools such as Redoc\footnote{\url{https://github.com/Redocly/redoc}} and SwaggerUI\footnote{\url{https://github.com/swagger-api/swagger-ui}} can automatically convert OpenAPI specifications into 
human-readable webpages, allowing developers to better understand these APIs. Additionally, specifications are commonly used in 
input validation~\cite{KarlssonQuickREST, FETOpenAPIValidation} and testing~\cite{morest, evomaster}. 

Common downstream clients of OpenAPI specifications leverage realistic examples of OpenAPI parameters (when they exist) as a part of their workflow. 
Fuzzers~\cite{arte, sourabhpaper} use examples to guide API testing, producing fewer invalid requests and covering deeper code paths.
Chat-bots~\cite{vaziri_et_al_2017,rizk_et_al_2020, schick_et_al_2023} first build an underlying model of a system and then
derive API calls from the natural language utterance. Recently-developed large language
models (LLMs), like ChatGPT~\cite{chatgpt} and GPT-4~\cite{openai2023gpt4},
benefit from using API parameter examples, and other LLMs use them for fine-tuning, as evaluated on dialog benchmarks~\cite{showdonttell}. API parameter examples can 
also improve human understanding, especially for novice users \cite{VenigallaHumanAPIUnderstanding, IchincoNoviceAPIExamples}.

However, despite their widespread adoption, most OpenAPI specifications lack API parameter examples (only 1,953 out of a dataset of 13,346 mined OpenAPI parameters have any examples). There has been some research on generating examples for OpenAPI specifications.
Prior work follows two approaches: (i)~extracting examples from API descriptions~\cite{sourabhpaper} or (ii)~mining examples from knowledge bases~\cite{arte, saigen}. The goal of both approaches is to generate diverse 
and correct examples. Example correctness is important, as these examples serve as input to software testing and dialog systems.
Conversely, example diversity is also important, as examples
  that differ from one another help testing increase its coverage and help
chat-bots generalize their natural-language understanding.
Both approaches to example generation are limited: mining examples only works for examples present in knowledge bases, while extracting examples from descriptions only works when the description explicitly enumerates parameter examples.

We present \emph{\technique}, which combines retrieval-based
prompting~\cite{retrievalprompting} with iterated calls to in-context
learning~(\textsc{Icl}) to generate diverse and correct API parameter examples.
\techniquecap leverages the ability of LLMs to generate realistic examples based on their pretraining. 
Unlike knowledge bases, LLMs are pretrained on large swaths of the internet, and thus have 
a strong prior of the world around them. We take as input the OpenAPI specification without examples and generate examples for all API parameters, regardless of whether examples exist on the internet or the descriptions specify example values. 
For correctness, we use greedy decoding (taking the highest probability token at each step) to generate one (likely) correct example. 
We perform postprocessing to only keep examples that are similar to our (likely)
correct example. 
For diversity, we both increase temperature, and, unlike vanilla
\textsc{Icl}, use iterated calls with multiple prompt contexts.
One can increase temperature (smoothing the distribution of next token probabilities) to generate different model outputs. 
Additionally, we observe that the problem of example diversity is similar to the challenge of generating different model outputs, which is solved by ensembles~\cite{ensembles} of different models. This observation
leads us to use multiple prompt contexts, where each context consists of a different set of few-shot examples.

We evaluate \technique, finding that it generates diverse, correctly typed examples. 
We further manually annotate a sample of 385 parameters and show that 75\% of the generated examples are correct. 
We then demonstrate the usefulness of the generated examples in three downstream settings: fuzzing, dialogue benchmarks, and
human API understanding, which we assess via an exploratory developer pilot.
Our examples significantly improve performance in these tasks, improving branch coverage by 116\%,  
dialog intent recognition by 3\%, and dialog slot filling by 5\%, compared to the original specifications.

To summarize, our core contributions are as follows:
\begin{itemize}
\item We identify adding examples as a \emph{single} improvement to API specifications that benefits \emph{several} downstream use cases (understanding, fuzzing, chat-bots).
\item Inspired by how ensembles use multiple models to improve results, we introduce \emph{\technique}, a new technique for using LLMs to generate API examples. We combine retrieval-based prompting, multiple prompt contexts, and post-processing to produce 
diverse yet correct examples.
\item We include an extensive experimental evaluation that quantifies the value of the generated examples for several use-cases. These include fuzz testing, chat-bots, and an exploratory study of developers' API understanding.
\end{itemize}

Our prompting, intrinsic, fuzzing, and exploratory study evaluation and code are at \url{https://figshare.com/s/8eec881ddf8e6573f43f}, including detailed reproduction instructions. 
We elide calls to internal company services in the prompting code, but release all other code. We are unfortunately unable to release our API parameter bank, intrinsic evaluation dataset, and SeqATIS dataset, as 
they are internal to the large technology company at which this work was conducted, but hope that the other elements of the artifact are informative for subsequent research.


\section{Motivating Example and Overview}

\begin{figure*}
    \centerline{\includegraphics[width=\textwidth]{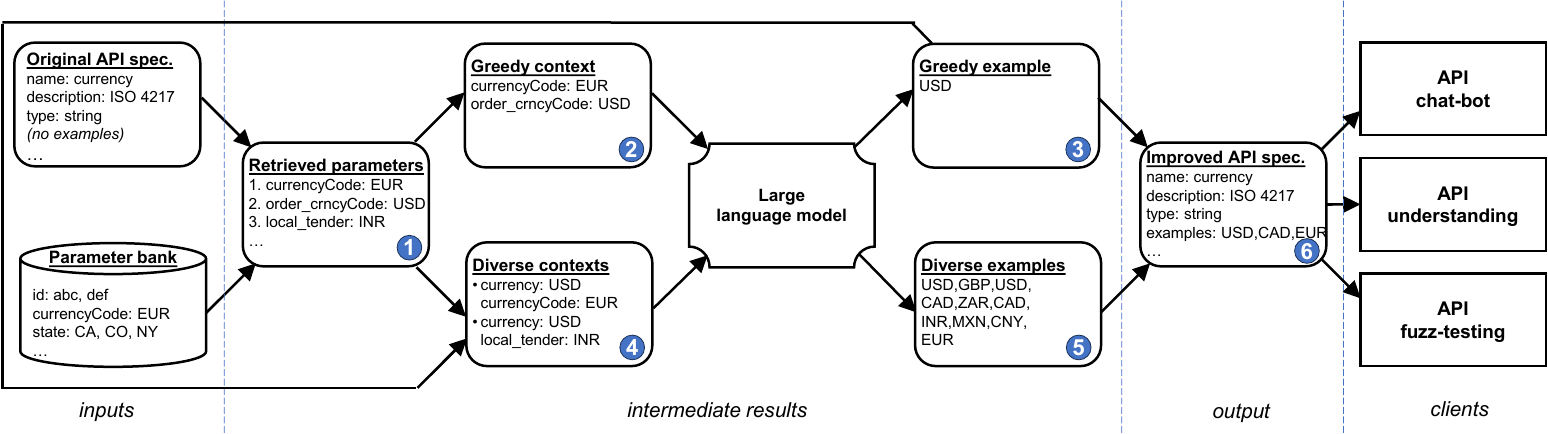}}
    \caption{\label{fig:overview}\small Overview and running example of our approach. Circled numbers correspond to different steps in our approach.}
\end{figure*}    

\begin{lstlisting}[float, language=yaml, caption={\small Illustrative OpenAPI parameter from the Rest Countries API. Prior approaches struggle to generate correct examples for this API parameter; 
    knowledge bases contain many false positives, and the description contains no examples.}, label=lst:exemplar_before]
name: currency
description: Search by ISO 4217 currency code
in: path
required: true
schema:
    type: string
\end{lstlisting}

\Cref{lst:exemplar_before} shows an illustrative parameter for the
\texttt{/currency}
endpoint\footnote{https://restcountries.com/v2/currency}
of the REST Countries API.
As with most OpenAPI parameters, this specification contains its name,
a short description, and a type.
However, it does not contain any example values, nor can example
values easily be extracted from the description or name.
To try the \texttt{/currency} endpoint, a developer would either need
domain knowledge of ISO 4217 currency codes or would need to search
for an example.
Fuzzers also fail to cover deeper code paths for this endpoint, as
they would start from a random sequence of bits and would only arrive
at a valid ISO 4217 currency code by chance.

Two common approaches to generating example values, namely mining them from a knowledge base such as DBPedia or extracting examples from the description, would also fail here. 
While ISO 4217 is an entity in DBPedia (the knowledge base used by the state-of-the-art example generation tool, ARTE~\cite{arte}), there are numerous other currency codes that are not ISO 4217, meaning 
that generated examples are semantically incorrect. ARTE~\cite{arte} circumvents this by calling the API with examples to see if they are valid; however, this limits applicability
to cases like fuzzing, which can send a large volume of requests to the API. The description also does not enumerate examples of currency codes that could be extracted.

\Cref{fig:overview} gives an overview of \technique. It first retrieves parameters from the API parameter bank that are similar to the parameter from the original API specification~(step~\circlednumber{1}). 
Then it creates a prompt context by greedily selecting the top-most similar retrieved parameters for in-context learning~(step~\circlednumber{2}). 
Following this, it uses the LLM with greedy decoding to obtain the greedy example, which has the highest confidence~(step~\circlednumber{3}). 
It then creates multiple diverse prompt contexts, each of which includes the greedy example plus some retrieved parameters for in-context learning, selected to be similar but with some randomization~(step~\circlednumber{4}),
and uses iterated calls to the LLM with a higher temperature to obtain multiple diverse examples, one from each of the diverse prompt contexts~(step~\circlednumber{5}). 
Then it creates a list of filtered examples that include the greedy example and some of the diverse examples, which it adds to the API specification~(step~\circlednumber{6}).



Our approach performs well on the snippet in \Cref{lst:exemplar_before}, generating \texttt{USD}, \texttt{CAD}, and \texttt{EUR}, all valid currency examples.

\section{\techniquecap}

\techniquecap takes an API specification without parameter examples as
input and returns an improved specification with those examples as output.
\Cref{fig:overview} outlines the approach.
Offline, we create a parameter bank by mining parameters and examples,
and pick an off-the-shelf LLM (\Cref{sec:model}).
Online, given a parameter in an OpenAPI specification, we retrieve relevant 
parameters from the parameter bank (\Cref{sec:retrievingrelevantparameters}),
build prompt contexts (\Cref{sec:promptcontextgeneration}),
and finally postprocess model output (\Cref{sec:postprocessing}).

\subsection{Offline: Mining Examples, Model Selection} 
\label{sec:miningapiexamples}
\label{sec:model}

We mine 1,236 OpenAPI specifications from API Guru~\cite{apis_guru} and Wittern et. al.~\cite{apiharmony}.
This collection has OpenAPI specifications for popular enterprise applications such as Box, Google Drive, YouTube, and others.
We parse the mined specifications to extract each API parameter and corresponding examples. Of 13,346 parameters, 1,953 have examples. 
We use these mined examples as our parameter bank (shown among the inputs on the left of \Cref{fig:overview}).

We use Falcon,\footnote{\url{https://huggingface.co/tiiuae/falcon-40b}} a 40B parameter model trained on one trillion tokens from the internet, for
the LLM (middle of \Cref{fig:overview}).
Falcon outperforms LLAMA, GPT-3, and MPT on the OpenLLM leaderboard. Falcon has also been extensively pretrained on code, which we hypothesize will help with
type correctness. 
Using a large but not huge open-source model such as Falcon is
representative of commercial settings that must balance cost and data
exposure regulatory concerns.
We model the task of example-generation as an instance of few-shot
prompting, varying the prompt context to generate different examples.

\subsection{Retrieving Relevant Parameters} 
\label{sec:retrievingrelevantparameters}

Given an API parameter, we first seek a set of relevant similar parameters from the parameter bank (\Cref{fig:overview}:~\circlednumber{1}).
We first extract the initial 50 characters from the API parameter description (our dataset has a median description length of 54 characters, with the first 50 characters concisely representing a parameter's purpose or function). 
At times, the full description is excessively verbose, with all 
other parameter information outside the description having a median length of 63 characters, thus truncating at 50 characters ensures that we do not overwhelm other important information.
We append the exact name of the API parameter to the description, ensuring that the name of the API parameter is factored into any similarity computation.
Lastly, we append the operation ID, which offers additional context about the operation associated with the parameter. For example, the parameter `\texttt{name}' has different meanings if the operation ID is `\texttt{getCountries}'
or `\texttt{getUserByUsername}'.
We use the concatenated string of the parameter description, parameter name, and operation ID as the \emph{query} for retrieval.

We use BM25~\cite{bm25} as the retrieval method, due to its high speed and accuracy~\cite{ahmed2023improving}.
BM25 calculates a weight of terms based on their frequency in both the query and the target documents. 
It then considers the term's prevalence across the entire parameter bank (intuitively infrequent terms discriminate better). 
When BM25 processes a parameter (such as `currency' in the running
example), it returns a similarity score for each API parameter in the
parameter bank.
These scores measure how closely each parameter in the
parameter bank matches the parameter we are generating examples for.
We leverage this distribution of similarity scores to craft prompt
\emph{contexts} (sets of few-shot examples for in-context learning).

\subsection{Prompt Context Generation}
\label{sec:promptcontextgeneration}

Prompt context generation consists of two phases: eliciting the greedy example, and then constructing ten
prompt contexts~(of five shots each) to elicit diverse examples.
We use a two-phase approach to improve both correctness and diversity of the generated examples.

The first phase prompts the LLM with the top five retrieved parameters with the highest similarity to the query (\Cref{fig:overview}:~\circlednumber{2}) as returned by retrieval (Section~\ref{sec:retrievingrelevantparameters}).
Greedy decoding in an LLM simply picks the most likely token at each generation step, thus deterministically yielding the sequence of most-probably tokens.
By leveraging greedy decoding, this step aims to produce a (likely) correct example~\circlednumber{3}.
 Our model yields \texttt{USD} as the greedy example (a correct currency code). By providing the greedy example in all prompt contexts, we ensure that the LLM, even at a higher temperature setting, 
 generates examples that align with the original example. 

The second phase improves example diversity by sampling from the distribution of similarity scores to generate 10 prompt 
contexts of five examples each~\circlednumber{4}.
We take inspiration from ensembles~\cite{ensembles}, where multiple models produce different outputs that improve both the correctness and diversity
of the resulting system. Our prompt contexts, each of which consists of a different set of API parameter examples (i.e., ``shots''), are similar to the diverse models used in ensembles.
We iteratively call the LLM with each prompt context with a higher temperature of 0.5 to generate 10 example candidates~\circlednumber{5}.
The order of these calls does not matter; they can be parallelized or batched.
In the running example, the calls return \texttt{USD}, \texttt{GPP}, \texttt{USD}, \texttt{CAD}, \texttt{ZAR}, \texttt{CAD}, \texttt{INR}, \texttt{MXN}, \texttt{CNY}, and \texttt{EUR}.

\begin{lstlisting}[float, language=yaml, xleftmargin=0pt, caption={\small LLM prompt for currency code. We provide the parameter that is missing examples and five few-shot examples.}, label=lst:llmprompt]
# Given an OpenAPI parameter, generate a unique example of the parameter.
input_0 = {
    "param_name": "currencyCode",
    "type": "string",
    "operation_id": "contractInfo",
    "description": "The currency code (ISO 4217)",
    "api_name": "beezup"
}
# must generate a unique currencyCode string
example_0 = "EUR" 
...
input_6 = {
    "param_name": "currency",
    "type": "string",
    "operation_id": "v2Currency",
    "description": "Search by ISO 4217 currency code",
    "api_name": "rest-countries"
}
# must generate a unique currency string
example_6 =
\end{lstlisting}

\subsection{Postprocessing}
\label{sec:postprocessing}

We perform postprocessing to narrow these 10 example candidates down to 3 examples to add to the improved API specification (\Cref{fig:overview}:~\circlednumber{6}).
First, we filter out all examples that do not match in type to the API parameter we are generating examples for.
This is the earliest opportunity for this filter, and we do it right away given the importance of type compatibility.
We then add the greedy example to the 10 example candidates and perform deduplication. We always
include the greedy example in our set of three generated examples, as it is likely to be correct. Following this, we add all examples that the model generates multiple times to our set of three, and return 
this set if it contains at least three examples. For example, if the model generates the currency \texttt{CAD} twice, then we add it to the final set of three examples.
If, at this point, there are fewer than three examples, we use BERT~\cite{devlin2018pretraining} to encode each example and the greedy example.
We then select the most similar examples until we have three
examples~(illustrated by adding \texttt{EUR} in \Cref{fig:overview}).

This ensures that the generated examples are similar in format and content to the greedy example, improving their likelihood of being correct.
We choose to favor correctness over diversity here, given its importance to downstream tasks~(testing and chat-bots). 
Using BERT embeddings  ensures that we are comparing the semantic similarity of each example to the greedy example, rather than doing a simple text-based match (which, in the case of currency codes, is less meaningful).

\section{Intrinsic Evaluation}
\label{sec:intrinsiceval}
While ultimately extrinsic evaluations (\Cref{sec:extrinsiceval})
matter most for downstream clients, they are laborious to measure, so
we used intrinsic evaluations for nimble iterative modeling.
We evaluate examples generated by \technique on intrinsic correctness
and diversity.
Specifically, we measure whether examples generated by \technique are
type correct, unique, and semantically diverse.
We also hand-evaluated a smaller subset of examples for semantic correctness.
We compare different components of \technique across these metrics.

\subsection{Experimental Setup}

\subsubsection{Dataset}

We evaluate modeling approaches on a randomly sampled dataset of 1,000 OpenAPI parameters mined from mainstream services including but not limited to Box, Google Drive, and Gmail. 
We remove all parameters in the parameter bank from our set of API parameters prior to sampling. 
We also remove all Boolean and enum parameters~(approximately 1,000 from the initial mined set) from our evaluation set, as predicting the values of these parameters is trivial. 
Due to computational cost, we do not run our intrinsic evaluation on the full final set of 13,346 parameters, instead focusing on a likely-representative random sample of 1,000 examples (approximately 1/13) of the dataset.
This sampling is in line with prior work~\cite{programFailuresEmpirical, GopinathSampleSize}, which sample a similar proportion of the dataset for evaluation.
These include 668~string, 129~array, 106~integer, 34~number, 14~object, and 5~datetime types. The remaining 44 parameters come from a variety of other types including color, tuples, and None types.

\subsubsection{Approach}

We evaluate the efficacy of each component of our approach (adding retrieval, sampling from the distribution of similarity scores, 
and applying our postprocessing).
This is equivalent to an ablation study: the final setting is the full approach, earlier settings remove components.
We prompt the model as described in each settings and evaluate the generated examples using the metrics described below.

\noindent\textbf{Static:} Static refers to a static prompt of five parameter and example pairings for in-context learning. We also include the greedy example as part of the prompt and use a temperature of 0.5. Temperature 
corresponds to the level of randomness in text generation - temperature of 0 refers to sampling the most likely tokens, while higher temperature refers to sampling more diversely. 
We prompt the LLM 10 times to generate 10 examples and perform deduplication. Finally, we randomly select three examples to return to the user.

\noindent\textbf{Retrieval:} Retrieval refers to the greedy retrieval approach. Rather than sampling 10 prompt contexts from the distribution of similarity 
scores, we only use a single prompt context containing the five most similar parameters for prompting.
In other words, this setting performs in-context learning with retrieval, but no iterated calls.

\noindent\textbf{Retrieval (w/contexts)}: Our retrieval with context approach adds iterated calls with context sampling.
Rather than selecting the five most similar examples for all 10 prompts, we 
build prompt contexts by randomly sampling from the distribution of similarity scores (similar parameters are more likely
to be chosen than different parameters).

\noindent\textbf{Retrieval (w/postprocessing)}: This is our final approach used in extrinsic evaluations (fuzzing, dialog, and exploratory usability study).
We apply our postprocessing that filters out type-incorrect examples and selects examples that are similar to the greedy example.
This helps ensure that our examples are correct, both in type and 
in semantic meaning (close to a generated example likely to be correct).

\subsubsection{Metrics}

We define the following set of metrics to benchmark various prompting approaches. The main factors we consider are example correctness and example diversity.

\noindent\textbf{Type Correctness:} Type correctness adheres to the strict definition of all generations from the 
LLM being the same type as the parameter. We use this strict definition to ensure all generations conform to the same example type.
Recall that our intrinsic evaluation focuses on open-ended types
(strings, numbers, arrays, objects, etc.) but not Boolean or enums
(as generating values for types with small closed sets is trivial).

\noindent\textbf{Uniqueness:} Uniqueness refers to the ability of the LLM to generate three case-insensitive unique examples from 10 generations.
Higher uniqueness values indicate more diverse LLM generated examples. For example, if all 10 generations are the same example, the uniqueness would be 
0, otherwise if there are three unique examples it would be 1.

\noindent\textbf{Diversity:} Diversity is 1 minus mean cosine similarity between the BERT~\cite{devlin2018pretraining} embeddings of 
examples. We choose to use BERT embeddings over TF-IDF or BM25 embeddings, as BERT embeddings detect semantic similarity, while other approaches only 
detect overlap of tokens (syntactic similarity).

\noindent\textbf{Example Correctness:} Example correctness refers to generated examples matching the specification. We define correctness as examples that both satisfy preconditions
specified in the natural language description of the parameter and have consistent format between all generated examples. Correct examples can be used in an API call to the 
API under test without 4xx or input validation errors.
Unlike the other metrics, which are fully automated, this metric requires human effort.
We manually annotate a randomly sampled subset of 385 out of our 1,000 sampled examples across all four settings, for 95\% confidence in the correctness results.

\subsection{Intrinsic Evaluation Results}

\begin{table}[h]
    \centering
    \caption{\small Intrinsic evaluation metrics on 1,000 (columns Type, Unique, Both (type correct and unique), Div) and 385 (column Correct) randomly sampled examples. Each approach component improves type correctness, the proportion 
    of unique examples, and overall correctness.}
    \label{tab:modelingbenchmarking}
    \begin{tabular}{l|r|r|r|r|r}
    \toprule
    \textbf{Setting} & \textbf{Type} & \textbf{Unique} & \textbf{Both} & \textbf{Div} & \textbf{Correct}\\
    \midrule
    static & 97\% & 48\% & 47\% & 0.22 & 70.4\% \\
    retrieval & 98\% & 55\% & 55\% & 0.20 & 73.2\% \\
    w/contexts & 98\% & 66\% & 65\% & \textbf{0.23} & 65.7\% \\
    w/postprocessing & \textbf{99\%} & \textbf{67\%} & \textbf{67\%} & 0.19 & \textbf{74.3\%} \\
    \bottomrule
    \end{tabular}
\end{table}

\Cref{tab:modelingbenchmarking} shows the results from running various components of \technique on a selected OpenAPI parameters.
We show how each component improves on the baseline.

We find that type correctness of generated examples is relatively strong across all approaches (varying from 97\% to 99\%). We hypothesize this is due
to LLMs' extensive training on code, where type is important in generating the next token. However, we do notice that type correctness does increase as we 
add retrieval-based prompting and our postprocessing, which improves type correctness to 99\%.

In terms of generating unique examples, we find that each step in our process improves upon the previous step. Retrieval and adding contexts see approximately a 10\%
improvement over the previous steps. Postprocessing improves uniqueness slightly, with 67\% of examples generated having 3 examples. Cosine similarity between 
examples remains relatively stable across modes, with retrieval templating (third row) slightly improving example diversity. We do want examples to have consistent format, while
still being diverse, likely resulting in lower diversity scores. The average Levenshtein edit distance on our dataset is 15 characters, suggesting the examples are still syntactically different from one another on average. 

Example correctness remains relatively stable across all four settings (varying from 66\% to 74\%).
The correctness of our final approach is higher
than any intermediate approach. Note that our evaluation of example correctness is conservative: in order for an example to be 
correct, all generations need to satisfy preconditions and have consistent format. Even examples not labeled as correct can 
still be useful for developers~(such as an example of a time parameter that is missing the timezone), meaning that the 74\% correctness rate is likely an underestimate of the true utility of the examples.
 Overall, our approach is often correct, showing the promise that LLMs pose for usefully enhancing API specifications. 
The extrinsic evaluation in the following section shows that, despite
not always being correct, synthetic examples benefit all three
downstream clients we tried.

\section{Extrinsic Evaluation}
\label{sec:extrinsiceval}

This section evaluates \technique on downstream tasks (clients on the right-hand side of \Cref{fig:overview}), namely software testing (\Cref{sec:softwaretesting}), API chatbots (\Cref{sec:chatbots}), and, by means of an exploratory pilot, human API understanding (\Cref{sec:apiunderstanding}).

\subsection{Software Testing}\label{sec:softwaretesting}

\begin{lstlisting}[float, language=yaml, caption={\small Fuzzing enhanced OpenAPI specification. Examples generated by \technique are both diverse and correct.}, label=lst:exemplar_after]
name: currency
description: Search by ISO 4217 currency code
required: true
schema:
    type: string
    enum:
    -   USD
    -   CAD
    -   EUR
example: USD
...
name: currency
description: Search by ISO 4217 currency code
required: true
schema:
    type: string
\end{lstlisting}

The goal of REST API testing is to find inputs that increase code coverage (and, ultimately, find bugs). 
At a high level, API fuzzers encode the schemas present in OpenAPI specifications, and use them to generate values for API endpoints. 
Coverage serves as a feedback mechanism: calls that increase coverage are saved for further mutation, while calls that do not are thrown out.

\subsubsection{Dataset and fuzzers} We evaluate \technique for fuzzing using a dataset from  Kim et. al.~\cite{sourabhpaper} consisting of both small and large APIs. We exclude the OMDB and Spotify APIs from that dataset, due to changes in both that make usage more challenging, and internal restrictions that block certain endpoints. 
This leaves seven widely-used REST API services --- FDIC, REST Countries, ohsome, GenomeNexus, OCVN, LanguageTool, and YouTube.  
This previous dataset included 4 that were run as local instances --- GenomeNexus, OCVN, LanguageTool, and YouTube --- for the purposes of computing coverage.  We therefore follow the previous evaluation and compute black-box performance on all~7 and coverage on the~4. 

We evaluate using four popular fuzzers: 
EvoMaster~\cite{evomaster}, Mo\-REST~\cite{morest}, RESTest~\cite{restest}, and RestTestGen~\cite{resttestgen}. We report results for each fuzzer along with the aggregated results across them all.
We used a version of RESTest that includes ARTE~\cite{arte}, a state-of-the-art example generation approach, as part of its implementation.
Hence, we refer to it as RESTest/ARTE below, helping show
how \technique compares to and can complement ARTE.

\subsubsection{Approach} To measure performance in the fuzzing context, we run \technique with each OpenAPI parameter that we extract from the fuzzing OpenAPI specifications. 
For each API parameter, we overload the specification with two options: the parameter with examples, and the parameter without examples, following Kim et al.~\cite{sourabhpaper}.
Since most fuzzers do not directly use API examples, we had to use a
work-around, where we encode the examples both using the example
attribute and as an enum.
\Cref{lst:exemplar_after} shows how we encode these values (the generated examples are \texttt{USD}, \texttt{CAD} and \texttt{EUR}). 
We also include the original parameter, to allow for fuzzers to explore values outside these examples.
This ensures the fuzzer can explore the example values and mutate existing example values 
by hitting the overloaded endpoint. For example, a fuzzer could choose the value \texttt{CAD} and then mutate it to \texttt{CDF} by hitting the overloaded endpoint twice. 
We run each fuzzer on both the original specification and the enhanced specification.

We were unfortunately unable to directly compare to the approach in Kim. et. al~\cite{sourabhpaper}.  We have filed an issue and have an ongoing discussion with the authors on the use of their artifact, and the paper does not directly ablate example generation.
We did randomly sample 700 of our 13,346 mined API parameters~(approximately~5\%), finding that only 43 enumerated examples occur in the description. Thus, even if Kim et. al.~\cite{sourabhpaper} extracts examples with 100\% accuracy, it could only do so for 6\% of all API parameters. 

\subsubsection{Metrics} We use a combination of API fuzzing metrics and code coverage to evaluate the performance change of adding examples to each fuzzer. 

\noindent\textbf{Proportion of 2xx Requests:} 2xx requests represent successful invocations of API endpoints, i.e., requests that yielded an HTTP response code between 200--299\footnote{\url{https://en.wikipedia.org/wiki/List_of_HTTP_status_codes}}.
These requests are saved for further fuzzing; thus having more 2xx 
requests means that the fuzzer is capable of testing functionality beyond simple input validation.

\noindent\textbf{Proportion of 4xx Requests:} 4xx requests represent poorly formatted invocations, where the fuzzer invokes the API incorrectly. Ideally, a fuzzer should make fewer 
4xx requests, as these are not testing deep functionality and wasting the fuzzing time budget.

\noindent\textbf{Proportion of 5xx Requests:} 5xx requests represent internal server errors. The goal of fuzzing is to catch such errors, thus more 5xx requests represent a successful fuzzing effort.

\noindent\textbf{Branch Coverage:} 
The goal of fuzzing efforts is to automatically test as much of the API as possible. Coverage is important, as higher code coverage indicates the fuzzer is testing a larger proportion of the API. We report branch coverage achieved by each fuzzer, as well as averaged across all four. 


\subsubsection{Results}
\begin{table*}[h]
    \caption{\small First three columns: API performance results for RESTest/ARTE, EvoMaster, MoREST, and RestTestGen across all 7 APIs. The proportion of 2xx requests goes up, 4xx goes down and 5xx slightly increases with enhanced examples. EvoMaster is the one exception, with
    2xx request proportions decreasing. Last column: Coverage results.  Coverage universally increases across all fuzzers with our enhancements. }
    \label{tab:fuzzing_results_runtime}
    \centerline{\small
    \begin{tabular}{@{}l|rrr|rrr|rrr|rrr@{}}
    \toprule
    & \multicolumn{3}{c|}{\textbf{Freq.\ of 2xx}} & \multicolumn{3}{c|}{\textbf{Freq.\ of 4xx}} & \multicolumn{3}{c|}{\textbf{Freq.\ of 5xx}} &  \multicolumn{3}{c}{\textbf{Branch Cov.}} \ \\
    \textbf{Tool Name} & \textbf{Base} & \textbf{Enhanced} & \textbf{Diff} & \textbf{Base} & \textbf{Enhanced} & \textbf{Diff} & \textbf{Base} & \textbf{Enhanced} & \textbf{Diff}  & \textbf{Base} & \textbf{Enhanced} & \textbf{Diff}  \\
    \midrule
    RESTest/ARTE & 0.28 & 0.31 & +11\% & 0.57 & 0.54 & -5\% & 0.10 & 0.10 & +0\% & 4.0 & 6.2 & +57\%  \\
    MoREST & 0.01 & 0.04 & +300\% & 0.83 & 0.81 & -2\% & 0.10 & 0.10 & +0\% & 3.2 & 17.5 & +447\% \\
    EvoMaster & 0.29 & 0.24 & -17\% & 0.58 & 0.62 & +7\% & 0.08 & 0.07 & -13\%  & 6.9 & 12.2 & +77\%\\
    RestTestGen & 0.21 & 0.26 & +24\% & 0.63 & 0.59 & -6\% & 0.10 & 0.13 & +30\% & 11.5 & 19.1 & +66\% \\
    \midrule
    \textbf{Average} & 0.20 & 0.21 & +5\% & 0.65 & 0.64 & -2\% & 0.10 & 0.10 & +0\% & 6.4 & 13.8 & +116\% \\
    \bottomrule
    \end{tabular}}
    \vspace{-4mm}
    \end{table*}   
 

\begin{lstlisting}[float, language=kotlin, belowskip=0pt, caption={\small An EvoMaster code snippet that explains why \technique does not improve proportion of 2xx requests. EvoMaster adds a default value of "EVOMASTER" to every enum, 
    causing our generated examples to degrade API fuzzing performance.}, label=lst:evomastercode]
if (schema.enum?.isNotEmpty() == true) {
    //Besides the defined values, add one to test robustness
    when (type) {
        "string" -> return EnumGene(name, (schema.enum as MutableList<String>).apply { add("EVOMASTER") })
    }
}
\end{lstlisting}

\Cref{tab:fuzzing_results_runtime} shows results across all fuzzers. 
Besides EvoMaster, all fuzzers exhibit a similar trend: examples lead to more 2xx requests~(around 3\%), fewer 4xx requests~(around 3\%), and around the same 5xx requests. 
This means that our example generation approach can seamlessly integrate with fuzzers to improve API testing, by better seeding fuzzers with realistic parameter 
examples that the fuzzers can use to invoke APIs. 

The reason EvoMaster does worse with \technique is their addition of a default value of EVOMASTER to each 
enum list.\footnote{\url{https://github.com/rapesil/EvoMaster/blob/master/core/src/main/kotlin/org/evomaster/core/problem/rest/RestActionBuilderV3.kt}}
\Cref{lst:evomastercode} shows the code snippet responsible for the performance degradation. This affects our approach because we use path overloading
to define an endpoint with our examples and a normal endpoint, allowing the fuzzer to perform standard mutation while leveraging our correct examples for 
better seeding. Due to this implementation detail, \technique leads to a higher proportion of 404 requests for EvoMaster.    

Branch coverage increases across all four fuzzers (from 6.4\% to 13.8\%). This means that in the time budget of one hour, more code is exercised.  
Although we do not show detailed results in the interest of space, method coverage also increases (from 11.9\% to 17.8\%, on average), meaning that more API enpoints are being hit as well. 

\subsection{API ChatBots}\label{sec:chatbots}
API chatbots take a natural-language utterance from a human and respond with the
appropriate set of API calls.
To accomplish this, API chatbots have intermediate tasks including recognizing which API a human wants to invoke~(intent recognition) 
and filling in the values for each API parameter~(slot filling). 
We evaluate how our examples affect the performance of these intermediate tasks on two benchmarks: MixATIS~\cite{mixatis} and Schema Guided Dialog (SGD)~\cite{sgd}.

\begin{lstlisting}[float, language=yaml, xleftmargin=0pt, caption={\small An example SeqATIS conversation and corresponding APIs. We use this model input and output to fine-tune an intent and slot filling chatbot on SeqATIS both with and without examples generated by \technique.}, label=lst:seqatis]
Model Input: what is the name of the airport in new york and then what is the distance between new york airport and downtown atlanta

Model Output: 
(1) API : "atis_airport",
    Parameters : [ city_name : philadelphia ]
(2) API : "atis_distance",
    Parameters : ...elided...
\end{lstlisting}

\begin{lstlisting}[float, style=conversation, belowskip=0pt, caption={\small SGD model input. We experiment with removing the example conversation and replacing the example values with examples generated by \technique.}, label=lst:sgd]
Example Conversation: [example]
[user] what's my balance? [system] in checking or savings?...
[slots] recipient_account_type=b of possible values a) checking b) savings

Model Input: [context]
[user] i'm paying some bills [system] which account? checking or savings?...
Model Output: [state] 
account_type=b of possible values a) checking b) savings...
\end{lstlisting}    

\subsubsection{Dataset}

MixATIS~\cite{mixatis} is a dialog benchmark developed to measure the ability of chatbots to deal with mixed-intent statements.
Mixed-intent statements consist of multiple API invocations in the same natural language 
statement. The goal is to successfully produce the correct set of APIs to invoke and fill in slot values for each API parameter. 

We use a version of the MixATIS dataset adapted for sequence to sequence (seq2seq) models called SeqATIS. Seq\-ATIS encodes all conversations and corresponding slot values in text.
\Cref{lst:seqatis} shows an example of a mixed turn conversation and the corresponding APIs. The input to the model is a sequence of text and the output from the model is a list of all APIs to call (intent recognition) and 
the values to send each API (slot filling).

Similarly, the SGD benchmark~\cite{sgd} is a diverse collection of dialogue interactions.
The benchmark covers multiple domains, including travel, services, and retail. 
Each conversation in the dataset is annotated with dialogue states and system actions. 

We follow the methodology of Gupta et. al.~\cite{showdonttell} in constructing our dataset. \Cref{lst:sgd} shows an example of how we~(and Gupta et. al.~\cite{showdonttell}) fine-tune LLMs on the SGD dataset. We show the model an example conversation and then prompt 
it with the current conversation. The model is fine-tuned to produce a list of slot values (values for each API parameter in the API being invoked).

The SeqATIS and SGD benchmarks are not well-formatted Open\-API specifications and lack attributes (e.g., operationId, endpoint path) required by ARTE, preventing a comparison against that technique here. \techniquecap is 
able to handle this case, allowing us to measure its impact on dialog tasks.

\subsubsection{Approach}
Since the SeqATIS dataset lacks descriptions, we use ChatGPT to generate descriptions for the 17 endpoints, and use these descriptions to generate examples with \technique. 
These examples are then used as a prefix to each conversation when fine-tuning the LLM. For these experiments, we use FlanT5-XXL~\cite{flant5}, as its instruction tuning enables it to generalize to 
 new unseen tasks. 

For our SGD evaluation, we investigate how our examples compare to the human-generated ``gold'' examples. We evaluate under the following settings: replacing the human-generated natural language
 conversation with just ``gold'' examples, replacing the conversation with examples generated by \technique, 
and not providing any examples.
For each of these settings, we fine-tune FlanT5-large.
Our baseline numbers are comparable with Gupta et. al~\cite{showdonttell}.

\subsubsection{Metrics} We evaluate the performance of dialog systems that are fine-tuned on the SeqATIS in both the intent recognition and slot filling tasks. 
For SeqAtis we measure both intent and slot filling exact match and normalized scores. Normalized scores, unlike exact match, allow for partial correctness, for example if 1/3 
slots in a natural-language utterance are matched correctly, the exact match slot score would be 0\% for that example while the normalized slot matches score would be 33\%. Overall score
measures correctness for both intent and slot matching for each API invoked in an utterance, and thus, is harder than either task individually. 

For SGD we only evaluate slot filling performance, as intents are provided as part of the input. We also report normalized slot match score, which is the proportion of matching slots across all data, rather 
than the stricter exact match score which requires all the slots in an example to be correct in order to be counted as correct.

\subsubsection{Results}

\begin{table}
    \centering
    \caption{\small SeqATIS intent recognition and slot filling scores. Original corresponds to the orignal specification, w/examples corresponds to adding examples generated by \technique and w/both corresponds to adding both API examples and API descriptions. 
    We find that examples generated by \technique along with descriptions generated by ChatGPT improve performance on both intent recognition and slot filling for SeqATIS.}
    \label{tab:seqatis}
    \begin{tabular}{l|rrr}
      \toprule
      \textbf{Metric} & \textbf{Original} & \textbf{w/examples} & \textbf{w/both} \\
      \midrule
      normalized intent & 0.92 & 0.94 & \textbf{0.95} \\
      normalized overall & 0.21 & 0.22 & \textbf{0.24} \\
      normalized slot matches & 0.76 & 0.76 & \textbf{0.81} \\
      exact match intent & 0.90 & 0.92 & \textbf{0.94} \\
      exact slot matches & 0.74 & 0.74 & \textbf{0.80} \\
      exact match overall & 0.20 & 0.21 & \textbf{0.23} \\
      \bottomrule
    \end{tabular}
  \end{table}

\Cref{tab:seqatis} shows results adding both examples and description as part of fine-tuning Flan-T5-XL.
Performance on 
both slot filling and intent detection improves substantially. Both examples and description improve overall performance, with 
normalized intent and overall score going up with examples. Slot matches stays the same with examples, but description
adds a significant performance improvement. Adding both examples and description improves performance between 3-5\%.

\begin{table}
    \centering
    \caption{\small SGD slot filling scores. We find that replacing human generated examples with synthetic examples generated by \technique results in little degradation in overall performance (1-2\% across all metrics).}
    \label{tab:sgd}
    \begin{tabular}{l|rr}
      \toprule
      \textbf{Setting} & \textbf{Exact slot} & \textbf{Normalized slot}\\
      \midrule
      hand-written examples & \textbf{0.74} & \textbf{0.95} \\
      no description & 0.73 & \textbf{0.95} \\
      \technique & 0.71 & 0.94\\
      no examples & 0.13 & 0.61\\
      \bottomrule
    \end{tabular}
    \vspace{-2mm}
  \end{table}

\Cref{tab:sgd} shows the performance degradation from replacing an example conversation with only example slot values and drop in performance from replacing 
example values with example values generated by \technique.
The core difference is for baseline and no-description settings, one would need to manually curate these 
example conversations or slot values, requiring developer time and effort.
Using \technique, we automatically generate slot examples for all parameters, 
with no human effort. The performance difference is not significant either, with an exact match slot score of 0.71 and normalized slot match score of 0.94 using \technique, while using gold examples has a 
exact match score of 0.73. Without examples, the performance on slot filling drops significantly, with an exact match score of 0.13 and normalized slot match score of 0.61. This shows that examples (even if they are 
synthetic) are essential for slot filling performance. For a 2\% degradation in exact match slot score and 1\% degradation in normalized slot match score, a developer could theoretically 
use a fully synthetic approach to generating examples.

\subsection{Human API Understanding}\label{sec:apiunderstanding}
\begin{lstlisting}[float, language=yaml, belowskip=0pt, caption={\small Genome-Nexus human study specification (with examples). We ask participants to write a natural language summary and to generate four valid cURL invocations of this API to assess API understanding.}, label=lst:humanstudy]
https://www.genomenexus.org/pfam/domain/{pfamAccession}:
    get:
        operationId: fetchPfamDomainsByAccessionGET
        parameters:
        -   name: pfamAccession
            description: A PFAM domain accession ID.
            required: true
            schema:
                type: string
                examples:
                -   PF00001
                -   PF00045
                -   PF00069
\end{lstlisting}

We conduct an exploratory study that aims to evaluate how useful the generated examples are for developers to understand a given OpenAPI Specification. 

\subsubsection{Dataset}
We select four API endpoints from our software testing dataset as specifications to evaluate human understanding of APIs. 
We choose APIs of varying difficulty.
Two are ``easy'', with a single parameter that is required, while two are ``moderate'', with six parameters, two of which are required. 
We modify specifications from the original to meet these requirements, for example marking parameters as required. 
We further modify specifications by removing any existing examples, any examples from the descriptions, and any default examples such that only generated examples are available to the participant. 
The modified endpoints used for the study are available as part of the replication kit.

\subsubsection{Approach}
We recruit six participants from a large technology company. 
  Due to the cost of recruiting expert participants, we had a limited sample size, however this is in line with Crasswell~\cite{Creswell2013}, 
  which states small samples can be suitable for certain exploratory studies.
Each participant is self-reported to be experienced in using OpenAPI Specifications. 
The study design is within-subjects~\cite{Creswell2013}, where each participant was exposed to two conditions: using a specification with and without generated examples. 
Each participant performed four tasks total (on four
  different APIs) across the two conditions,
  shuffled to avoid cross-condition learning effects.
Each condition has a pair of easy and moderate difficulty endpoints with the same two tasks per endpoint: 
write a summary of what the endpoint does in natural language and create four example invocations of the endpoint. \Cref{lst:humanstudy} shows an example of an easy specification. There is only one parameter marked as required. This specification has three examples generated by \technique. 
We shuffle the order of conditions and endpoints used per participant such 
that half of the participants start without examples and that endpoints are evenly distributed between conditions.
As we request example invocations using cURL (a simple command-line tool for transferring data with URLs),\footnote{https://curl.se/} we also provide a brief tutorial for writing cURL invocations that is available for each participant. 
At the end of the four tasks, we ask participants to provide open-ended feedback for how they understand API specifications, examples in specifications, and the study in general. Two researchers administered each study and recorded detailed notes. Although we did not perform a formal qualitative analysis for this exploratory study, 
we provide anecdotes from participants and notes to give some nuance to the quantitative results.

\subsubsection{Metrics}
In addition to qualitative metrics, we compute both accuracy and task completion time. For accuracy, we evaluate if the participant's generated cURL queries would return a 200 status code. We 
consider cURL queries where the parameters are correct values but the syntax is incorrect as correct for the purposes of our study. However, if any of the API parameters has any values that are incorrect, we mark the entire query as incorrect. We also 
measure the time to complete each task for both settings. We report both the average time to complete each task and accuracy for each combination of task and setting. 

\subsubsection{Results}

\begin{table}
    \centering
    \caption{\small Summary of tasks and settings. We find that providing developers with examples allows for faster task completion rates, with no effect on overall correctness.}
    \label{tab:tasks}
    \begin{tabular}{llrrrr}
        \toprule
        \textbf{API} & \textbf{Examples} & \textbf{Time (s)} & \textbf{Correct} & \textbf{Attempted} \\
        \midrule
        \multirow{2}{*}{Task 1} & Yes & 414 & 9/12 & 10/12 \\
        & No & 712 & 6/12 & 11/12 \\
        \midrule
        \multirow{2}{*}{Task 2} & Yes & 168 & 6/12 & 12/12 \\
        & No & 289 & 6/12 & 11/12 \\
        \midrule
        \multirow{2}{*}{Task 3} & Yes & 664 & 7/12 & 9/12 \\
        & No & 708 & 11/12 & 12/12 \\
        \midrule
        \multirow{2}{*}{Task 4} & Yes & 620 & 5/8 & 8/8 \\
        & No & 468 & 4/12 & 10/12 \\
        \bottomrule
    \end{tabular}
    \vspace{-2mm}
\end{table}

\Cref{tab:tasks} shows the quantitative results from our exploratory study of six experts. 
The average task completion time decreased when generated examples were present (around 50\%). This aligns with multiple participants' 
qualitative statements, where participants report using examples when understanding the meaning of an API.
 One surprising case was the decrease in accuracy for Task 3 from 11/12 to 7/12 when examples were present. This is likely due to the particpant attempting fewer of the 
  problems, along with some of the generated examples being incorrect and blindly copied.

Interestingly, for more difficult tasks, the time effect decreased, with 
1/2 intermediate tasks taking longer with examples than without examples. This could be because of bloated specifications; one participant stated ``But I generally tell my developers, 
get your parameter examples out of my code because you're just adding too much junk.'' Too many examples can pollute the specification, making it harder
to read and comprehend for a human. Accuracy remains relatively stable across both conditions, with participants eventually figuring out how to solve the task with and without API examples (just taking more time
to do so).

In tasks where no examples were provided, participants repeatedly complained about the overall quality of specifications. One participant stated, 
``I think it's a pretty bad API'', while another stated ``Sample IDs would be useful.'' Without examples, developers struggle to understand and 
use APIs in their workflows.

In the majority of cases, participants seemed to appreciate having generated examples over not having any examples. Multiple participants
reported improved guessing based off of examples, with one stating ``I can only guess based on the examples'' and another stating ``At least this one is good, 
it has examples.'' One limitation of these LLM-generated examples is that participants tended to blindly trust the API examples provided to them, with the majority
of participants in our study at some point copying an example into the cURL request. Since LLM-generated examples are not guaranteed to be 
fully accurate, blindly trusting these examples as oracles can lead to an increased proportion of badly formed or incorrect requests.

\section{Discussion}

We discusses the broader impacts of our work in the context of software documentation, chat bots, and fuzz testing.


\techniquecap complements existing automated documentation approaches: generated API examples can
be used to enhance existing API documentation. \techniquecap can generate examples for any API, making \technique more useful 
for company-specific APIs.  \techniquecap works as APIs evolve, while prior approaches would rely on
 the changed API existing on the internet or values being enumerated in the description.

\techniquecap improves the performance of chat-bots by providing LLM-generated examples of internal APIs, giving them context to better understand the inputs and outputs of each API. This leads to both improved intent recognition (which API to call) and slot filling 
(what parameter values to pass). This also applies to customer-facing APIs, with LLM-generated examples both improving documentation and automated chat-bot services. Our evaluation of \technique on 
both MixATIS and SGD datasets shows that our examples improve API performance across both intent recognition and slot filling. Interestingly, we find that LLM-generated examples achieve comparable performance to human-generated examples, with only a 2\% drop 
on slot filling exact match on SGD. 

Another important area for companies is testing~\cite{PetrovicMutationGoogle, machalica2019predictive}. Even though developers write unit and integration tests, many corporate services continue 
to be under-tested. Fuzzing helps find bugs in APIs~\cite{AFLplusplus-Woot20, ossfuzz}, 
and has been widely adopted by companies such as Google~\cite{googlefuzzing}.
\techniquecap improves the performance of fuzzers by providing realistic examples of API parameters, which can be used to seed these fuzzers, scaling to large APIs such as YouTube. Examples generated using \technique improve 
both API coverage and the proportion of 2xx requests.
Practically, this means that companies can add \technique to their
workflows and explore deeper code paths within the same fuzzing
budget.

\section{Limitations and Threats}
\noindent\textbf{Internal Threats:} An internal threat to validity is our implementation of \technique. 
To mitigate this, we used well-known programming libraries to construct prompts and publicly release our code. 
Another concern is that the LLMs we used may have seen API parameters at pretraining time.
We attempt to mitigate by using Falcon 40B, a large and new model (which generally have lower leakage rates than older and smaller models)~\cite{ramos2024largelanguagemodelsmemorizing}.

\noindent\textbf{External Threats:} An external threat is that we do not evaluate new bugs found by our technique for our fuzzing evaluation. 
Accurately bucketing fuzz crashes is an open challenge and we lack access to source code for all tested endpoints (instead we follow the metrics in NLP2Rest~\cite{sourabhpaper} and the original ARTE paper~\cite{arte}).
Finally, the small sample size of our exploratory study of six expert participants could limit the broader applicability of our findings. 
As such, the outcomes from our exploratory study should be considered as preliminary and interpreted with caution until larger, more comprehensive studies can be conducted. 

\noindent\textbf{Construct Threats:} A construct validity concern is our selection of evaluation metrics. 
We used metrics commonly employed in evaluating dialog systems, fuzzing, and the broader field of machine learning, including coverage and exact match. 
While widely accepted, these metrics may not fully capture the complexities of the dialog systems we are analyzing. 

\section{Related Work}

\noindent\textbf{OpenAPI Specification Enhancement:} Several approaches aim to create components of 
OpenAPI specifications from natural language specifications~\cite{huo_et_al_2022, yang_et_al_2018, baudart_et_al_2020_automl_icml}.
ARTE~\cite{arte} mines examples from knowledge bases for use in fuzz testing.
Recent approaches~\cite{sourabhpaper, decrop2024restnowautomatedspecification} use language models to generate 
requests and other OpenAPI fields.
Unlike existing techniques, which require 
examples to occur in knowledge bases or Open\-API descriptions, \techniquecap works on all OpenAPI specifications, generating relevant parameter examples.

\noindent\textbf{Leveraging OpenAPI Specifications:} There has been extensive work 
leveraging OpenAPI specifications for a variety of downstream tasks including chatbots, intent recognition, and business workflows~\cite{vaziri_et_al_2017,babkin_et_al_2017,rizk_et_al_2020}. Multiple fuzzers~\cite{godefroid_huang_polishchuk_2020, restler, evomaster} 
actively use refined specifications provided by OpenAPI to improve API testing. Unlike these prior works, which focused on a single downstream task each,
we evaluate our approach across several downstream tasks.

\noindent\textbf{Large Language Models:} Large language models (LLMs) can perform well across many tasks when prompted with 
instructions and examples~\cite{BrownGPT3, touvron2023llama}. LLMs such as ChatGPT~\cite{chatgpt}, GPT-4~\cite{openai2023gpt4} and 
LLAMA~\cite{touvron2023llama} perform well on a large range of natural language~\cite{naturalquestions, triviaqa} and code~\cite{humaneval, llmexampleshuman, nashid_sintaha_mesbah_2023} and testing~\cite{tan_et_al_2012} tasks with 
minimal examples.

We leverage these LLMs to automatically improve OpenAPI specifications. 
We chose to not use extremely large models such as GPT-4 or Copilot due to
company data leakage concerns and their high operating costs.
We did experiment with several smaller models than Falcon-40B.
While the results with these models were worse in absolute terms,
\technique yielded similar improvement on these models in relative terms.


\section{Conclusion}

We developed \technique, an LLM-based prompting approach to generate examples for OpenAPI parameters, performing both an intrinsic and multiple extrinsic 
evaluations. Intrinsically, \technique is capable of generating both correct and diverse examples.
Extrinsically, \technique's 
examples can be applied to a wide variety of downstream tasks, including software testing and dialog systems.
Our generated examples significantly improve performance in these tasks, increasing branch coverage by 116\%,  
dialog intent recognition by 3\%, and dialog slot filling by 5\%, compared to the original specifications.
Overall, \technique leverages the strong prior of LLMs to generate Open\-API examples for a wide variety of API parameters that were not possible in prior work, improving the downstream 
performance of several tasks that use API examples.

\balance

\bibliographystyle{IEEEtran}
\bibliography{bibliography}

\balance

\end{document}